\documentclass[letterpaper,twocolumn,10pt]{article}
\usepackage{usenix,epsfig,endnotes}

\usepackage[utf8]{inputenc}
\usepackage{color}
\usepackage{cite}
\usepackage{url}
\usepackage{graphicx}
\usepackage{authblk}
\usepackage{xspace}
\usepackage{array}
\usepackage{mathtools}
\usepackage{amsmath}
\usepackage{amssymb}
\usepackage{subcaption}
\usepackage{booktabs}
\usepackage{fancyhdr}

\newif\ifsubmit
\submitfalse

\ifsubmit
\newcommand{\dawn}[1]{}
\newcommand{\bo}[1]{}
\newcommand{\kevin}[1]{}
\newcommand{\ef}[1]{}
\newcommand{\ivan}[1]{}
\newcommand{\amir}[1]{}
\newcommand{\yoshi}[1]{}
\newcommand{\wh}[1]{}
\newcommand{\flo}[1]{}
\else
\newcommand{\dawn}[1]{\textcolor{red}{Dawn: #1}}
\newcommand{\bo}[1]{\textcolor{blue}{Bo: #1}}
\newcommand{\kevin}[1]{\textcolor{cyan}{Kevin: #1}}
\newcommand{\ef}[1]{\textcolor{red}{(Earl: #1)}}
\newcommand{\ivan}[1]{\textcolor{green}{Ivan: #1}}
\newcommand{\amir}[1]{\textcolor{cyan}{Amir: #1}}
\newcommand{\yoshi}[1]{\textcolor{blue}{Yoshi: #1}}
\newcommand{\wh}[1]{\textcolor{red}{wh: #1}}
\newcommand{\flo}[1]{\textcolor{red}{Florian: #1}}
\fi

\newcommand{\ignore}[1]{}
\newcommand{\algfull}[0]{Robust Physical Perturbations\xspace}
\newcommand{\algshort}[0]{RP\textsubscript{2}\xspace}
\newcommand{\ie}[0]{\emph{i.e.,}\xspace}
\newcommand{\etal}[0]{\emph{et al.}\xspace}
\newcommand{\eg}[0]{\emph{e.g.,}\xspace}

\begin{document}

\date{}

\pagestyle{fancy}
\chead{Extended version of our paper that appeared at WOOT 2018, co-located with USENIX Security.}
\rhead{}
\renewcommand{\headrulewidth}{0pt}

\title{\Large \bf Physical Adversarial Examples for Object Detectors}

\author{
{\rm Kevin Eykholt\textsuperscript{1}, Ivan Evtimov\textsuperscript{2}, Earlence Fernandes\textsuperscript{2}, Bo Li\textsuperscript{3}, \\Amir Rahmati\textsuperscript{4,6}, Florian Tram{\`e}r\textsuperscript{5}, Atul Prakash\textsuperscript{1}, Tadayoshi Kohno\textsuperscript{2}, Dawn Song\textsuperscript{3}}\\
\textsuperscript{1}University of Michigan\\
\textsuperscript{2}University of Washington\\ 
\textsuperscript{3}University of California, Berkeley\\
\textsuperscript{4}Stony Brook University\\
\textsuperscript{5}Stanford University\\
\textsuperscript{6}Samsung Research America
} 


\maketitle
\thispagestyle{fancy}
\pagestyle{empty}

\begin{abstract}

Deep neural networks (DNNs) are vulnerable to \emph{adversarial examples}---maliciously crafted inputs that cause DNNs to make incorrect predictions. Recent work has shown that these attacks generalize to the physical domain, to create perturbations on physical objects that fool image classifiers under a variety of real-world conditions. 
Such attacks pose a risk to deep learning models used in safety-critical cyber-physical systems.

In this work, we extend physical attacks to more challenging \emph{object detection models}, a broader class of deep learning algorithms widely used to detect and label multiple objects within a scene.
Improving upon a previous physical attack on image classifiers, we create perturbed physical objects that are either ignored or mislabeled by object detection models. We implement a \emph{Disappearance Attack}, in which we cause a Stop sign to ``disappear'' according to the detector---either by covering the sign with an adversarial Stop sign poster, or by adding adversarial stickers onto the sign.
In a video recorded in a controlled lab environment, the state-of-the-art YOLO v2 detector failed to recognize these adversarial Stop signs in over 85\% of the video frames. In an outdoor experiment, YOLO was fooled by the poster and sticker attacks in 72.5\% and 63.5\% of the video frames respectively. We also use Faster R-CNN, a different object detection model, to demonstrate the \emph{transferability} of our adversarial perturbations. The created poster perturbation is able to fool Faster R-CNN  in 85.9\% of the video frames in a controlled lab environment, and 40.2\% of the video frames in an outdoor environment. Finally, we present preliminary results with a new \emph{Creation Attack}, wherein innocuous physical stickers fool a model into detecting nonexistent objects.



\end{abstract}

\section{Introduction}

Deep neural networks (DNNs) are widely applied in computer vision, natural language, and robotics, especially in safety-critical tasks such as autonomous driving~\cite{lillicrap2015continuous}. At the same time, DNNs have been shown to be vulnerable to \emph{adversarial examples}~\cite{goodfellow2014explaining,papernot2016limitations,carlini2017towards,sabour2015adversarial,kos2017adversarial}, maliciously perturbed inputs that cause DNNs to produce incorrect predictions. These attacks pose a risk to the use of deep learning in safety- and security-critical decisions. For example, an attacker can add perturbations, which are negligible to humans, to a Stop sign and cause a DNN embedded in an autonomous vehicle to misclassify or ignore the sign.





Early works studied adversarial examples in the digital space only. However, it has recently been shown that it is also possible to create perturbations that survive under various physical conditions (\eg object distance, pose, lighting, etc.)~\cite{rp2,sharif2016accessorize,kurakin2016adversarial,athalye2017synthesizing, brown2017adversarial}. These works focus on attacking \emph{classification networks}, \ie models that produce a single prediction on a static input image.
In this work, we start exploring physical adversarial examples for \emph{object detection networks}, a richer class of deep learning algorithms that can detect and label multiple objects in a scene. Object detection networks are a popular tool for tasks that require real-time and dynamic recognition of surrounding objects, autonomous driving being a canonical application. Object detectors are known to be vulnerable to digital attacks~\cite{digitaldetectorattacks}, but their vulnerability to physical attacks remains an open question.

Compared to classifiers, object detection networks are more challenging to attack: 1) Detectors process an entire scene instead of a single localized object. This allows detectors to use contextual information (\eg the orientation and relative position of objects in the scene) to generate predictions. 2) Detectors are not limited to producing a single prediction. Instead, they label every recognized object in a scene, usually by combining predictions of the \emph{location} of objects in a scene, and of the labeling of these objects. Attacks on object detectors need to take both types of predictions (presence/absence of an object and nature of the object) into account, whereas attacks on classifiers only focus on modifying the label of a single (presumably present) object.

To create proof-of-concept attacks for object detectors, we start from the existing \algfull (\algshort) algorithm~\cite{rp2} of Eykholt \etal, which was originally proposed to produce robust physical attacks on image classifiers. The approach taken by Eykholt \etal (as well as by others~\cite{athalye2017synthesizing, kurakin2016adversarial}) is to sample from a distribution that mimics physical perturbations of an object (\eg view distance and angle), and find a perturbation that maximizes the probability of mis-classification under this distribution.
We find that the physical perturbations considered in their work are insufficient to extend to object detectors. 

Indeed, when working with image classifiers, prior works considered target objects that make up a large portion of the image and whose relative position in the image varies little. 
Yet, when performing object detection in a dynamic environment such as a driving car, the relative size and position of the multiple objects in a scene can change drastically. These changes produce additional constraints that have to be taken into account to produce successful robust physical attacks.
Many object detectors, for instance, split a scene into a grid or use a sliding window to identify regions of interest, and produce separate object predictions for each region of interest. As the relative position of an object changes, the grid cells the object is contained in (and the corresponding network weights) change as well. Robust perturbations, thus, have to be applicable to multiple grid cells simultaneously. 
We show that robustness to these physical modifications can be attained by extending the distribution of inputs considered by Eykholt \etal to account for additional synthetic transformations to objects in a scene (\eg changes in perspective, size, and position).

Following Eykholt \etal, we consider physical adversarial attacks on the detection and classification of Stop signs, an illustrative example for the safety implications of a successful attack. The perturbations, while large enough to be visible to the human eye, are constrained to resemble human-made graffiti or subtle lighting artifacts that could be considered benign. 
We consider an untargeted attack specific to object detectors, which we refer to as a \emph{Disappearance Attack}. In a Disappearance Attack, we create either an adversarial poster or physical stickers applied to a real Stop sign (see Figure~\ref{fig:noise}), which causes the sign to be ignored by an object detector in different scenes with varying object distance, location, and perspective. This attack is analogous to the one considered by Eykholt \etal for image classifiers, but targets a richer class of deep neural networks.

We further introduce a new \emph{Creation Attack}, wherein physical stickers that humans would ignore as being inconspicuous can cause an object detector into recognizing nonexistent Stop signs. This attack differs from prior attacks that attempt to fool a network into mis-classifying one object into another, in that it creates an entirely new object classification. 
Specifically, we experiment with creating adversarial stickers (similar to the ones considered in~\cite{brown2017adversarial}). Such stickers could for instance be used to mount \emph{Denial of Service}  attacks on road-sign detectors.

For our experiments, we target the state-of-the-art YOLO v2 (You Only Look Once) object detector~\cite{yolo}. YOLO v2 is a deep convolutional neural network that performs real-time object detection for 80 object classes. Our indoor (laboratory) and outdoor experiments show that up to distances of 30 feet from the target object, detectors can be tricked into {\em not} perceiving the attacker's target object using poster and sticker perturbations.\\

\noindent \textbf{Our Contributions:}
\begin{itemize}
    \item We extend the \algshort algorithm of Eykholt \etal to provide proof-of-concept attacks for object detection networks, a richer class of DNNs than image classifiers.
    
    \item Using our new and improved algorithm, we propose a new physical attack on object detection networks: the Disappearance Attack that cause physical objects to be ignored by a detector.
    
    \item We evaluate our attacks on the YOLO v2 object detector in an indoor laboratory setting and an outdoor setting. Our results show that our adversarial poster perturbation fools YOLO v2 in 85.6\% of the video frames recorded in an indoor lab environment and in 72.5\% of the video frames recorded in an outdoor environment. Our adversarial stickers fool YOLO v2 in 85\% of the video frames recorded in a laboratory environment and in 63.5\% of the video frames recorded in an outdoor environment.

    \item We evaluate the transferability of our attacks using the Faster R-CNN object detector in laboratory and outdoor environments. Our results show that our attacks fool Faster R-CNN in 85.9\% of the video frames recorded in a laboratory environment and in 40.2\% of the video frames recorded in an outdoor environment.
    
    \item We propose and experiment with a new type of \emph{Creation} attack, that aims at fooling a detector into recognizing adversarial stickers as non-existing objects. Our results with this attack type are preliminary yet encouraging.
\end{itemize}

Our work demonstrates that physical perturbations are effective against object detectors, and leaves open some future questions: 1) Generalization to other physical settings (\eg moving vehicles, or even real autonomous vehicles). 2) Further exploration of other classes of attacks: Our work introduces the disappearance and creation attacks which use posters or stickers, yet there are other plausible attack types (\eg manufacturing physical objects that are not recognizable to humans, but are recognized by DNNs). 3) Physical attacks on segmentation networks. We envision that future work will build on the findings presented here, and will create attacks that generalize across physical settings (\eg real autonomous vehicles), and across classes of object detection networks (\eg semantic segmentation~\cite{digitaldetectorattacks}).



\section{Related Work}
Adversarial examples for deep learning were first introduced by Szegedy \etal~\cite{szegedy2014intriguing}. Since their seminal work, there have been several works proposing more efficient algorithms for generating adversarial examples \cite{papernot2016limitations,goodfellow2014explaining,carlini2017towards,moosavi2015deepfool}. All of these works assume that the attacker has ``digital-level'' access to an input, \eg that the attacker can make arbitrary pixel-level changes to an input image of a classifier. For uses of deep learning in cyber-physical systems (\eg in an autonomous vehicle), these attacks thus implicitly assume that the adversary controls a DNN's input system (\eg a camera).
A stronger and more realistic threat model would assume that the attacker only controls the physical layer, \eg the environment or objects that the system interacts with, but not the internal sensors and data pipelines of the system.

This stronger threat model was first explored by Kurakin \etal They generated physical adversarial examples by printing digital adversarial examples on paper~\cite{kurakin2016adversarial}. In their work, they found that a significant portion of the printed adversarial examples fooled an image classifier. However, their experiments were done without any variation in the physical conditions such as different viewing angles or distances. 

Athalye \etal improved upon the work of Kurakin \etal by creating adversarial objects that are robust to variations in viewing angle~\cite{athalye2017synthesizing}. To account for such variations, they model small scale transformations synthetically when generating adversarial perturbations. They demonstrate several examples of adversarial objects that fool their target classifiers, but it is not clear how many transformations their attack is robust to. In their paper, they state their algorithm is robust to rotations, translations, and noise and suggest their algorithm is robust so long as the transformation can be modeled synthetically. 

Eykholt \etal also proposed an attack algorithm capable of generating physical adversarial examples~\cite{rp2}. Unlike Athalye \etal, they choose to model image transformations both synthetically and physically. Certain image transformations, such as changes in viewing angle and distance, are captured in their victim dataset. They apply other image transformations, such as lighting, synthetically when generating adversarial examples. Their work suggests that sole reliance on synthetic transformations can miss subtleties in the physical environment, thus resulting in a less robust attack. Different from all prior work that focused on {\em classifiers}, our work focuses on the broader class of object detection models. Specifically, we extend the algorithm of Eykholt \etal using synthetic transformations (perspective, position, scale) to attack object detection models.  


Lu \etal performed experiments using adversarial road signs printed on paper with the YOLO object detector~\cite{noneed}. Their results suggested that it is very challenging to fool YOLO with physical adversarial examples. 

Concurrent to our work, Chen \etal attacked the Faster R-CNN object detector \cite{DBLP:journals/corr/abs-1804-05810}. Their attack relies on generating adversarial poster perturbations that replace the road sign to fool Faster R-CNN. Our work differs in that we introduce two different attacks on object detectors, the disappearance and creation attacks, which use either adversarial poster or sticker perturbations. We also show black-box transferability from YOLO to the Faster-RCNN detector.

Our work resolves the challenges and
shows that existing algorithms can be adapted to produce physical attacks on object detectors in highly variable environmental conditions. 


\section{Background on Object Detectors}
Object classification is a standard task in computer vision. Given an input image and a set of class labels, the classification algorithm outputs the most probable label (or a probability distribution over all labels) for the image.
Object classifiers are limited to categorizing a single object per image. If an image contains multiple objects, the classifier only outputs the class of the most dominant object in the scene. In contrast, object detectors both locate and classify multiple objects in a given scene.

The first proposed deep neural network for object detection was Overfeat~\cite{overfeat}, which combined a sliding window algorithm and convolution neural networks. A more recent proposal, Regions with Convolutional Neural Networks (R-CNN) uses a search algorithm to generate region proposals, and a CNN to label each region. A downside of R-CNN is that the region proposal algorithm is too slow to be run in real-time. Subsequent works---Fast R-CNN~\cite{fast} and Faster R-CNN~\cite{ren15fasterrcnn}---replace this inefficient algorithm with a more efficient CNN. 

The above algorithms treat object detection as a two-stage problem consisting of region proposals followed by classifications for each of these regions. In contrast, so-called ``single shot detectors'' such as YOLO~\cite{redmon2016you} (and the subsequent YOLO v2~\cite{yolo}) or SSD~\cite{liu2016ssd} run a single CNN over the input image to jointly produce confidence scores for object localization and classification. As a result, these networks can achieve the same accuracy while processing images much faster. In this work, we focus on YOLO v2, a state-of-the-art object detector with real-time detection capabilities and high accuracy.

\begin{figure}[t]
  \center
  \includegraphics[width=0.9\textwidth]{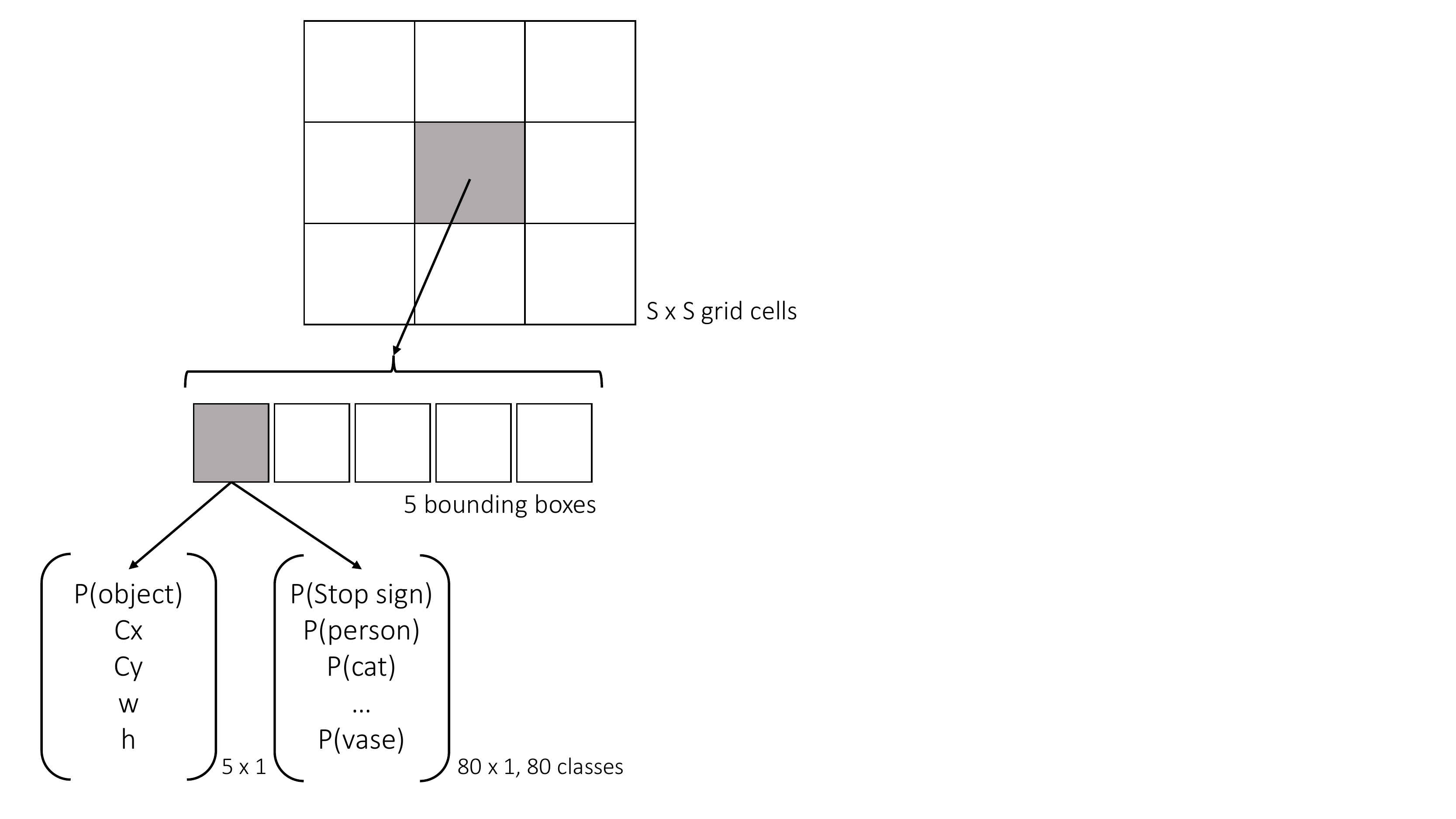}
  \caption{For an input scene, the YOLO v2 CNN outputs a $19 \times 19 \times 425$ tensor. To generate this tensor, YOLO divides the input image into a square grid of $S^2$ cells ($S = 19$). For each grid cell, there are B bounding boxes ($B = 5$). Each bounding box predicts 5 values: probability of an object in the cell, co-ordinates of the bounding box (center x, center y, width, height). Additionally, for each bounding box the model predicts a probability distribution over all 80 output classes.}
  \label{fig:yolo_output}
\end{figure}

The classification approach of YOLO v2 is illustrated in Figure~\ref{fig:yolo_output}. A single CNN is run on the full input image and predicts object location (bounding boxes) and label confidences for $361$ separate grid cells (organized into a $19 \times 19$ square over the original image). For each cell, YOLO v2 makes a prediction for $5$ different boxes. For each box, the prediction contains the box confidence (the probability that this box contains an object), its location and the probability of each class label for that box. A box is discarded if the product of the box confidence and the probability of the most likely class is below some threshold (this threshold is set to $0.1$ in our experiments). Finally, the \emph{non-max suppression} algorithm is applied in a post-processing phase to discard redundant boxes with high overlap~\cite{yolo}.

Such an object detection pipeline introduces several new challenges regarding physical adversarial examples: First, unlike classification where an object is always assumed present and the attack only needs to modify the class probabilities, attacks on a detector network need to control a combination of box confidences and class probabilities for all boxes in all grid cells of the input scene. Second, classifiers assume the object of interest is centered in the input image, whereas detectors can find objects at arbitrary positions in a scene. Finally, the object's size in the detector's input is not fixed. In classification, the image is usually cropped and resized to focus on the object being classified. Object detectors are meant to reliably detect objects at multiple scales, distances and angles in a scene.

These challenges mainly stem from object detectors being much more flexible and broadly applicable than standard image classifiers. Thus, albeit harder to attack, object detectors also represent a far more interesting attack target than image classifiers, as their extra flexibility makes them a far better candidate for use in reliable cyber-physical systems.



\section{Physical Adversarial Examples for Object Detectors}
We will first summarize the original \algshort algorithm, before discussing the modifications necessary to adapt the algorithm to attack object detectors.

\subsection{The \algshort Algorithm}
The \algshort algorithm proposed by Eykholt \etal optimizes the following objective function:

\begin{equation}
\begin{multlined}
\underset{\delta}{\mathrm{argmin}}~\lambda ||M_{x} \cdot \delta||_{p} + \mathit{NPS}(M_{x} \cdot \delta)\\ +  \mathbb{E}_{x_i \sim X^V} J(f_{\theta}(x_i + T_i( M_{x} \cdot \delta)), y^{*})
\end{multlined}
\label{eq:original}
\end{equation}

\begin{figure}[t]
  \centering
  \includegraphics[width=0.35\textwidth]{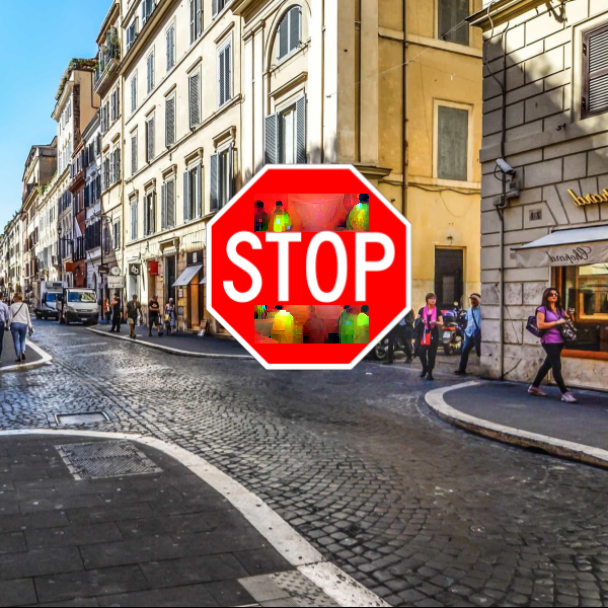}
  \caption{An example of an adversarial perturbation overlaid on a synthetic background. The Stop sign in the image is printed such that it is the same size as a U.S. Stop sign. Then, we cut out the two rectangle bars, and use the original print as a stencil to position the cutouts on a real Stop sign.}
  \label{fig:noise}
\end{figure}

The first term of the objective function is the $\ell_{p}$ norm (with scaling factor $\lambda$) of the perturbation $\delta$ masked by $M_{x}$. The mask is responsible for spatially constraining the perturbation $\delta$ to the surface of the target object. For example, in Figure \ref{fig:noise}, the mask shape is two horizontal bars on the sign. 

\begin{figure}[t]
    \centering
    \begin{subfigure}[b]{0.3\textwidth}
        \includegraphics[width=\textwidth]{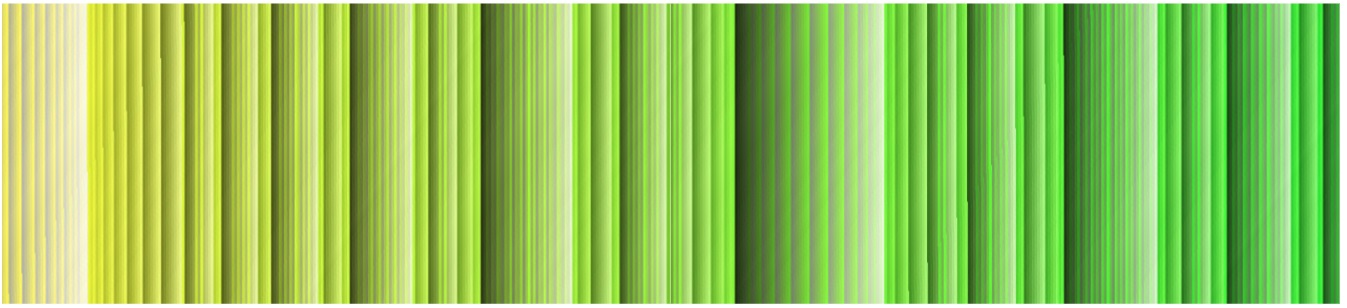}
        \caption{Digital Image}
        \label{fig:originalcolor}
    \end{subfigure}
    ~ 
    \begin{subfigure}[b]{0.3\textwidth}
        \includegraphics[width=\textwidth]{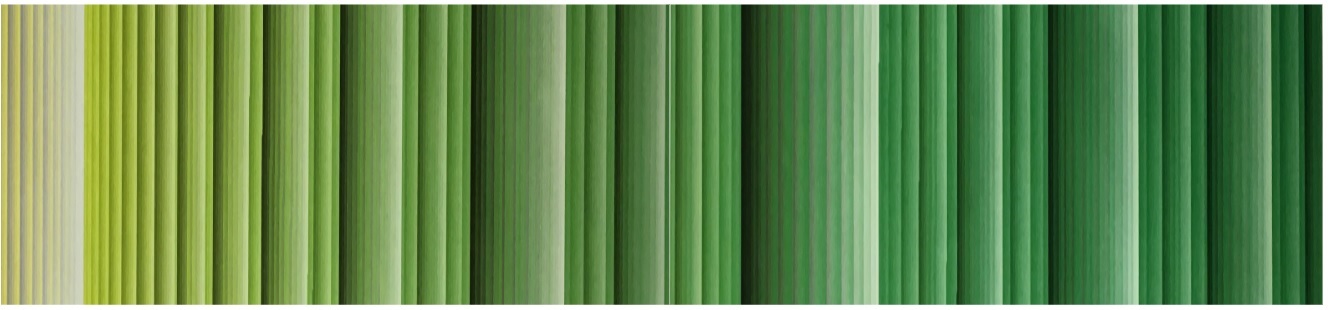}
        \caption{Printer Result of Digital Image}
        \label{fig:printedcolor}
    \end{subfigure}
    \caption{The image in (a) shows the image as it is stored digitally. The result of printing and taking a picture of the image in (a) is shown in (b).}
    \label{fig:colorerror}
\end{figure}

The second term of the objective function measures the printability of an adversarial perturbation. Eykholt \etal borrow this term from prior work \cite{sharif2016accessorize}. The printability of a perturbation is affected by two factors. First, the colors the computed perturbation must reproduce. Modern printers have a limited color gamut, thus certain colors that appear digitally may not be printable. Second, a printer may not faithfully reproduce a color as it is shown digitally (see Figure \ref{fig:colorerror}).

The last term of the objective function is the value of the loss function, $J(\cdot,\cdot)$ averaged across all of the images sampled from $X^V$. In practice, this is a set of victim images. The victim dataset is composed of multiple images of the object taken under a variety of physical conditions such as changes in viewing angle, viewing distance and lighting. $T_{i}$ is an ``alignment function'' that applies a digital transformation that mimics the physical conditions of victim object $x_i$. For example, if the victim object $x_i$ is a rotated version of the ``canonical'' target object, then the perturbation $M_x \cdot \delta$ should also be rotated appropriately. 
Thus, to simulate physical consistency of the perturbed object, we apply the alignment function $T_i$ to the masked perturbation. $f_\theta(\cdot)$ is the output of the classifier network, and $y^{*}$ is the adversarial target class.

\subsection{Extensions to \algshort for Object Detectors}
Our modified version of \algshort contains three key differences from the original algorithm proposed by Eykholt \etal First, due to differences in the output behavior of classifiers and object detectors, we make modifications to the adversarial loss function. Second, we observed additional constraints that an adversarial perturbation must be robust to and model these constraints synthetically. Finally, we introduce a smoothness constraint into the objective, rather than using the $\ell_{p}$ norm. In the following, we discuss each of these changes in detail.

\subsubsection{Modified Adversarial Loss Function}
An object detector outputs a set of bounding boxes and the likelihood of the most probable object contained within that box given a certain confidence threshold. See Figure~\ref{fig:yolo_output} for a visualization of this output. By contrast, a classifier outputs a single vector where each entry represents the probability that the object in the image is of that type. Attacks on image classifiers typically make use of the cross-entropy loss between this output vector, and a one-hot representation of the adversarial target. However, this loss function is not applicable to object detectors due to their richer output structure. Thus, we introduce a new adversarial loss function suitable for use with detectors. This loss function is tailored to the specific attacks we introduce in this work.

\paragraph{Disappearance Attack Loss.} The goal of the attacker is to prevent the object detector from detecting the target object. To achieve this, the adversarial perturbation must ensure that the likelihood of the target object in any bounding box is less than the detection threshold (the default is 25\% for YOLO v2). In our implementation of the attack, we used the following loss function:

\begin{equation}
\begin{multlined}
$$
J_{d}(x, y) = \max_{s \in S^2, b \in  B} P (s, b, y, f_{\theta}(x))
$$
\end{multlined}
\label{eq:masking}
\end{equation}

Where $f_{\theta}(x)$ represents the output of the object detector (for YOLO v2, this is a $19 \times 19 \times 425$ tensor). $P(\cdot)$ is a function that extracts the probability of an object class from this tensor, with label $y$ (in our case, this is a Stop sign) in grid cell $s$ and bounding box $b$. We denote $x$ as the input scene containing our perturbed target object. 

Therefore, the loss function outputs the maximum probability of a Stop sign if it occurs within the scene. Using this loss function, the goal of the adversary is to directly minimize that probability until it falls below the detection threshold of the network.

\paragraph{Creation Attack Loss.}
We propose a new type of \emph{Creation Attack}, wherein the goal is to fool the model into recognizing nonexistent objects. Similar to the ``adversarial patch'' approach of~\cite{brown2017adversarial}, our goal is to create a physical sticker that can be added to any existing scene. Contrary to prior work, rather than causing a mis-classification our aim is to create a new classification (\ie a new object detection) where non existed before.

For this, we use a composite loss function, that first aims at creating a new object localization, followed by a targeted ``mis-classification.'' The mask $M_x$ is sampled randomly so that the adversarial patch is applied to an arbitrary location in the scene.
As above, let $f_{\theta}(x)$ represent the full output tensor of YOLO v2 on input scene $x$, and let $P(s,b,y,f_{\theta}(x))$ represent the probability assigned to class $y$ in box $b$ of grid cell $s$. Further let $P_{\text{box}}(s,b,f_{\theta}(x))$ represent the probability of the box only, \ie the model's confidence that the box contains \emph{any} object. Our loss is then
\begin{align}
\textrm{object} &= P_{\text{box}}(s,b,f_{\theta}(x)) > \tau \nonumber\\
J_{c}(x, y) &= \textrm{object} + (1-\textrm{object}) \cdot P(s,b,y,f_{\theta}(x)) 
\label{eq:create}
\end{align}

Here, $\tau$ is a threshold on the box confidence (set to $0.2$ in our experiments), after which we stop optimizing the box confidence and focus on increasing the probability of the targeted class. As our YOLO v2 implementation uses a threshold of $0.1$ on the product of the box confidence and class probability, any box with a confidence above $0.2$ and a target class probability above $50\%$ is retained.

\subsubsection{Synthetic Representation of New Physical Constraints}
Generating physical adversarial examples for detectors requires simulating a larger set of varying physical conditions than what is needed to trick classifiers. In our initial experiments, we observed that the generated perturbations would fail if the object was moved from its original position in the image. This is likely because a detector has access to more contextual information when generating predictions. As an object's position and size can vary greatly depending on the viewer's location, perturbations must account for these additional constraints.

To generate physical adversarial perturbations that are positionally invariant, we chose to synthetically model two environmental conditions: object rotation (in the Z plane) and position (in the X-Y plane). In each epoch of the optimization, we randomly place and rotate the object. Our approach differs from the original approach used by Eykholt \etal, in that they modeled an object's rotation physically using a diverse dataset. We avoided this approach because of the added complexity necessary for the alignment function, $T_{i}$, to properly position the adversarial perturbation on the sign. Since these transformations are done synthetically, the alignment function, $T_{i}$, simply needs to use the same process to transform the adversarial perturbation.

\subsubsection{Noise Smoothing using Total Variation}
The unmodified \algshort algorithm uses the $\ell_{p}$ norm to smooth the perturbation. However, in our initial experiments, we observed that the $\ell_{p}$ norm results in very pixelated perturbations. The pixelation hurts the success rate of the attack, especially as the distance between the viewer and the object increases. We found that using the total variation norm in place of the $\ell_{p}$ norm gave smoother perturbations, thus increasing the effective range of the attack. Given a mask, $M_{x}$, and noise $\delta$, the total variation norm of the adversarial perturbation, $M_{x} \cdot \delta$, is:

\begin{equation}
\begin{multlined}
TV(M_{x} \cdot \delta) = \\ \sum_{i,j} |(M_{x} \cdot \delta)_{i+1,j} - (M_{x} \cdot \delta)_{i,j}|\\ + |(M_{x} \cdot \delta)_{i,j+1} - (M_{x} \cdot \delta)_{i,j}|
\end{multlined}
\label{eq:totalvariation}
\end{equation}

where $i,j$ are the row and column indices for the adversarial perturbation. Thus our final modified objective function is:

\begin{equation}
\begin{multlined}
\underset{\delta}{\mathrm{argmin}}~\lambda TV(M_{x} \cdot \delta) + \mathit{NPS}\\ +  \mathbb{E}_{x_i \sim X^V} J_d(x_i + T_i( M_{x} \cdot \delta),y^{*})
\end{multlined}
\label{eq:modified}
\end{equation}

where $J_d(\cdot,y^{*})$ is the loss function (discussed earlier) that measures the maximum probability of an object with the label $y^{*}$ contained in the image. In our attack, $y^*$ is a Stop sign.

\section{Evaluation}
\begin{figure}[t]
    \centering
    \includegraphics[width=0.45\columnwidth]{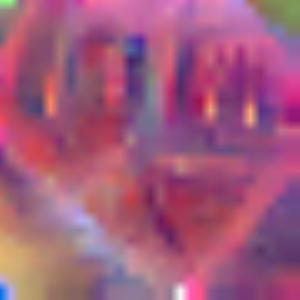}
    \caption{Patch created by the Creation Attack, aimed at fooling YOLO v2 into detecting nonexistent Stop signs.}
    \label{fig:patch}
\end{figure}

We first discuss our experimental method, where we evaluate attacks in a whitebox manner using YOLO v2, and in a blackbox manner using Faster-RCNN. Then, we discuss our results, showing that state-of-the-art object detectors can be attacked using physical posters and stickers. Figure~\ref{fig:highresattacks} shows the digital versions of posters and stickers used for disappearance attacks, while Figure~\ref{fig:patch} shows a digital version of the sticker used in a creation attack.

\begin{figure}[t]
    \includegraphics[width=0.45\columnwidth]{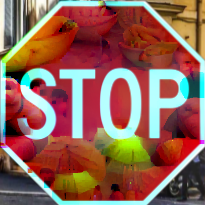}
    \includegraphics[width=0.45\columnwidth]{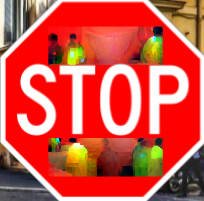}
    \caption{Output of the extended \algshort algorithm to attack YOLO v2 using poster and sticker attacks.}
    \label{fig:highresattacks}
\end{figure}

\ignore{
\begin{figure}[t]
    \centering
    
    \begin{subfigure}{\columnwidth}
        \center
        \includegraphics[width=0.5\textwidth]{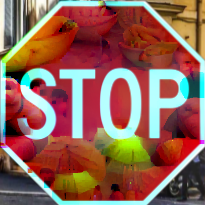}
        \caption{Adversarial Poster Perturbation}
        \label{fig:octagon}
    \end{subfigure}
    \begin{subfigure}{\columnwidth}
        \center
        \includegraphics[width=0.5\textwidth]{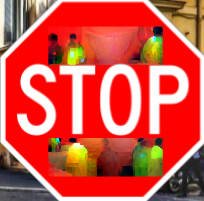}
        \caption{Adversarial Sticker Perturbation}
        \label{fig:orig}
    \end{subfigure}
    
    ~ 
    \caption{Output of the extended \algshort algorithm to attack YOLO v2 using poster and sticker attacks.}
    \label{fig:highresattacks}
\end{figure}
}

\subsection{Experimental Setup}
We evaluated our disappearance attack in a mix of lab and outdoor settings. For both the poster and sticker attacks, we generated adversarial perturbations and recorded several seconds of video. In each experiment, recording began 30 feet from the sign and ended when no part of the sign was in the camera's field of view. Then, we fed the video into the object detection network for analysis. We used the YOLO v2 object detector as a white-box attack. We also ran the same videos through the Faster-RCNN network to measure black-box transferability of our attack.

For the creation attack, we experimented with placing stickers on large flat objects (\eg a wall or cupboard), and recording videos within 10 feet of the sticker.


\subsection{Experimental Results}

We evaluated the perturbations for a disappearance attack using two different masks and attacked a Stop sign.  First, we tested a poster perturbation, which used an octagonal mask to allow adversarial noise to to be added anywhere on the surface of the Stop sign. Next, we tested a sticker perturbation. We used the mask to create two rectangular stickers positioned at the top and bottom of the sign. The results of our attack are shown in Table \ref{tab:disappearance_result}. 

\begin{table}[t]
    \centering
    \begin{tabular}{c c c}
    \toprule
       YOLO v2  & Poster & Sticker  \\ \toprule
    Indoors     & 202/236 (85.6\%)    & 210/247 (85.0\%)      \\ \midrule
    Outdoors    & 156/215 (72.5\%)    & 146/230 (63.5\%)      \\ \bottomrule
    \end{tabular}
    \caption{Attack success rate for the disappearance attack on YOLO v2. We tested a poster perturbation, where a true-sized print is overlaid on a real Stop sign, and a sticker attack, where the perturbation is two rectangles stuck to the surface of the sign. The table cells show the ratio: number of frames in which a Stop sign was {\em not} detected / total number of frames, and a success rate, which is the result of this ratio.}
    \label{tab:disappearance_result}
\end{table}

\begin{figure*}[t]
    \centering
    
    \begin{subfigure}[b]{\textwidth}
        \includegraphics[width=\textwidth]{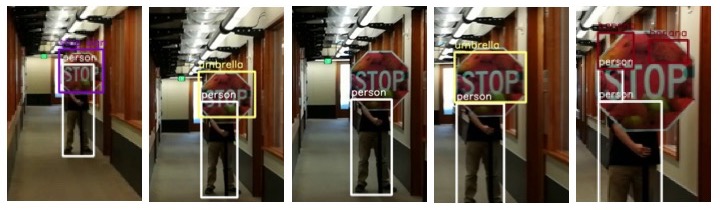}
        \caption{The poster attack inside}
        \label{fig:poster-inside-sequence}
    \end{subfigure}
    \begin{subfigure}[b]{\textwidth}
        \includegraphics[width=\textwidth]{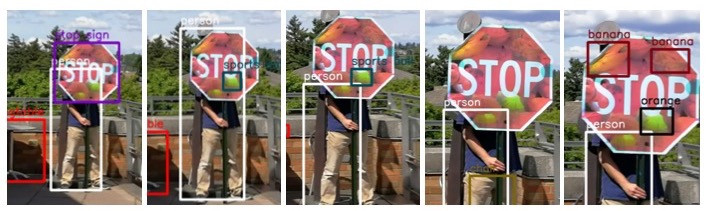}
        \caption{The poster attack outside}
        \label{fig:poster-outside-sequence}
    \end{subfigure}
    
    \begin{subfigure}[b]{\textwidth}
        \includegraphics[width=\textwidth]{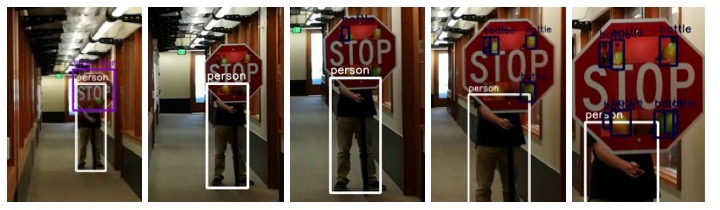}
        \caption{The sticker attack inside}
        \label{fig:sticker-inside-sequence}
    \end{subfigure}
    
    \begin{subfigure}[b]{\textwidth}
        \includegraphics[width=\textwidth]{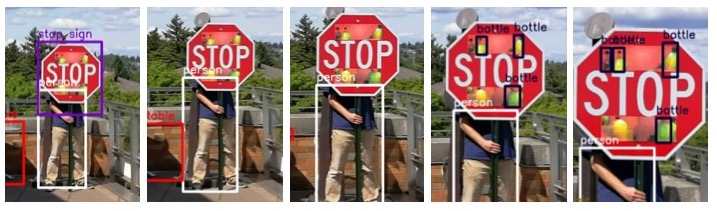}
        \caption{The sticker attack outside}
        \label{fig:sticker-outside-sequence}
    \end{subfigure}
    ~ 
    \caption{Sample frames from our attack videos after being processed by YOLO v2. In the majority of frames, the detector fails to recognize the Stop sign.}
    \label{fig:sequence}
\end{figure*}

In indoor lab settings, where the environment is relatively stable, both the poster and sticker perturbation demonstrate a high success rate in which at least 85\% of the total video frames do not contain a Stop sign bounding box. When we evaluated our perturbations in an outdoor environment, we notice a drop in success rate for both attacks. The sticker perturbation also appears to be slightly weaker. We noticed that the sticker perturbation did especially poorly when only a portion of the sign was in the camera's field of view. Namely, when the sticker perturbation began to leave the camera's field of view, the Stop sign bounding boxes appear very frequently. In contrast, this behavior was not observed in the poster perturbation experiments, likely because some part of the adversarial noise is always present in the video due to the mask's shape. Figure \ref{fig:sequence} shows some frame captures of our adversarial Stop sign videos.

To measure the transferability of our attack, we also evaluated the recorded videos using the Faster R-CNN object detection network.\footnote{We used the Tensorflow-Python implementation of Faster R-CNN found at \url{https://github.com/endernewton/tf-faster-rcnn} It has a default detection threshold of 80\%}. The results for these experiments are shown in Table \ref{tab:rcnn_results}. 

\begin{table}[tb]
    \centering
    \begin{tabular}{c c c}
    \toprule
       FR-CNN  & Poster & Sticker  \\ \toprule
    Indoors     & 189/220 (85.9\%)    & 146/248 (58.9\%)      \\ \midrule
    Outdoors    & 84/209 (40.2\%)    & 47/249 (18.9\%)      \\ \bottomrule
    \end{tabular}
    \caption{Attack success rate for the disappearance attack on Faster R-CNN. We tested a poster perturbation, where the entire Stop sign is replaced with a true-sized print, and a sticker attack, where the perturbation is two rectangles stuck to the surface of the sign. The table cells show the ratio: number of frames in which a Stop sign was {\em not} detected / total number of frames, and a success rate, which is the result of this ratio.}
    \label{tab:rcnn_results}
\end{table}

We see from these results that both perturbations transfer with a relatively high success rate in indoor lab settings where the environment conditions are stable. However, once outdoors, the success rate for both perturbations decreases significantly, but both perturbations retain moderate success rates. We observe that our improved attack algorithm can generate an adversarial poster perturbation, which transfers to other object detection frameworks, especially in stable environments.

\begin{figure}[!t]
    \centering
    
    \includegraphics[width=0.4\textwidth]{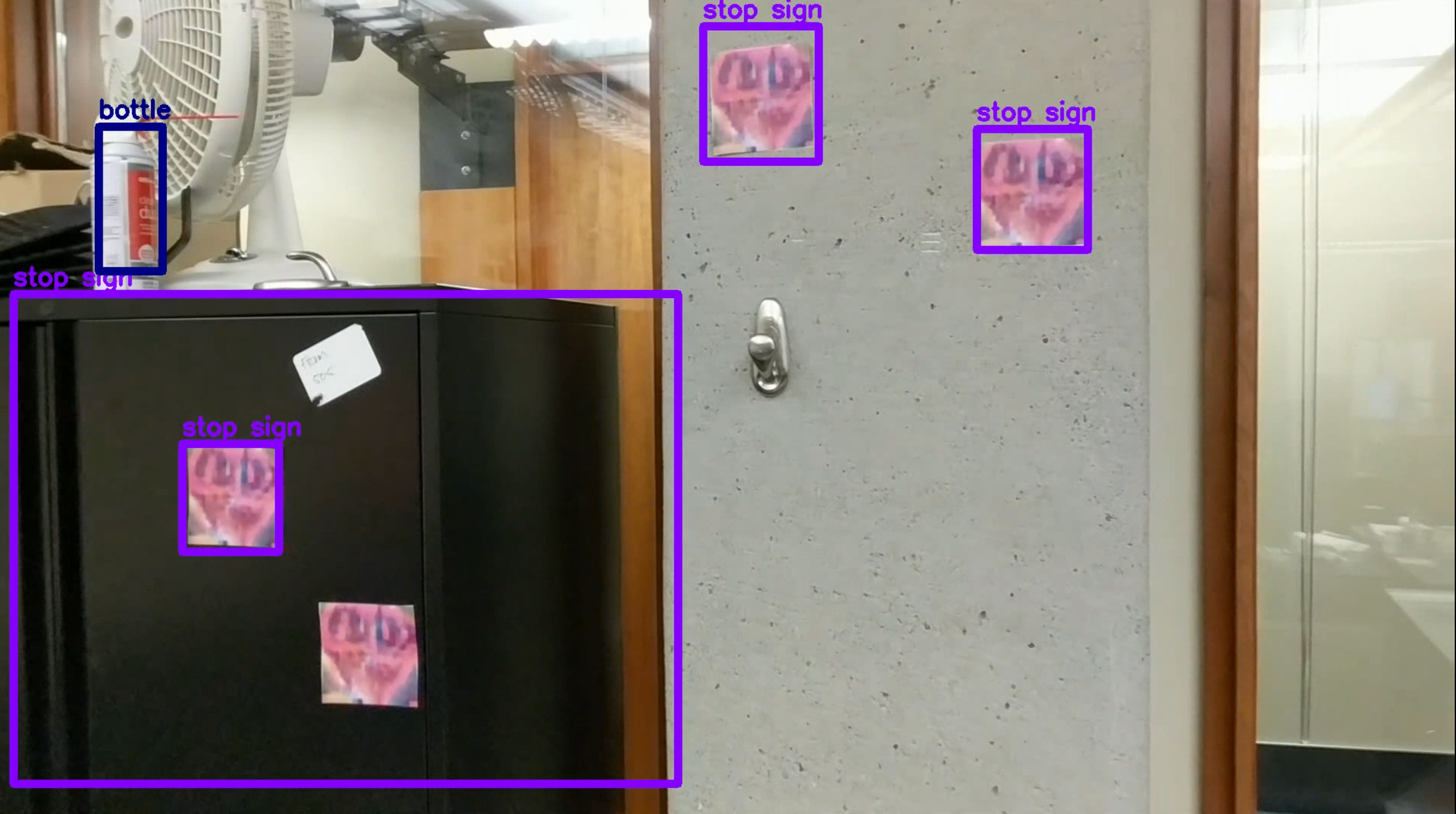}
    \caption{Sample frame from our creation attack video after being processed by YOLO v2. The scene includes 4 adversarial stickers reliably recognized as Stop signs.}
    \label{fig:patch_frame}
\end{figure}

Finally, we report on some preliminary results for creation attacks (the results are considered preliminary in that we have spent considerably less time optimizing these attacks compared to the disappearance attacks---it is thus likely that they can be further improved). When applying multiple copies of the sticker in Figure~\ref{fig:patch} to a cupboard and office wall, YOLO v2 detects stop signs in $25\%$--$79\%$ of the frames over multiple independent videos. A sample video frame is shown in Figure~\ref{fig:patch_frame}. Compared to the disappearance attack, the creation attack is more sensitive to the sticker's size, surroundings, and camera movement in the video. This results in highly variable success rates and is presumably because (due to resource constraints) we applied fewer physical and digital transformations when generating the attack. Enhancing the reliability and robustness of our creation attack is an interesting avenue for future work, as it presents a novel attack vector (\eg DOS style attacks) for adversarial examples.

\section{Discussion}
In the process of generating physical adversarial examples for object detectors, we note several open research questions that we leave to future work.

\paragraph{Lack of detail due to environmental conditions.} We noticed physical conditions (\eg poor lighting, far distance, sharp angles), which only allowed macro features of the sign (\ie shape, general color, lettering) to be observed clearly. Due to such conditions, the details of the perturbations were lost, causing it to fail. This is expected as our attack relies on the camera being able to perceive the adversarial perturbations somewhat accurately. When extreme environmental conditions prevent the camera from observing finer details of the perturbation on the sign, the adversarial noise is lost. We theorize that in order to successfully fool object detectors under these extreme conditions, the macro features of the sign need to be attacked. For example, we could create attachments on the outside edges of the sign in order to change its perceived shape.

\paragraph{Alternative attacks on object detectors.} In this work, we explored attacking the object detector such that it fails to locate an object,
or that it detects non-existent objects.
There are several alternative forms of attack we could consider. One alternative is to attempt to generate physical perturbations that preserve the bounding box of an object, but alter its label (this is similar to targeted attacks for classifiers). Another option is to generate further 2D or even 3D objects that appear nonsensical to a human, but are detected and labeled by the object detector. The success of either of these attacks, which have been shown to work digitally \cite{digitaldetectorattacks,noiseclassifierattacks}, would have major safety implications.

\paragraph{Extensions to semantic segmentation.} A broader task than object detection is semantic segmentation---where the network labels every pixel in a scene as belonging to an object. Recent work has shown digital attacks against semantic segmentation~\cite{digitaldetectorattacks}. An important future work question is how to extend current attack techniques for classifiers, and detectors (as this work shows) to create physical attacks on segmentation networks.

\paragraph{Impact on Real Systems.} Existing cyber-physical systems such as cars and drones integrate object detectors into a control pipeline that consists of pre- and post-processing steps. The attacks we show only target the object detection component in isolation (specifically YOLO v2). Understanding whether these attacks are capable of compromising a full control pipeline in an end-to-end manner is an important open question. Although YOLO v2 does recognize a Stop sign in some frames from our attack videos, a real system would generally base its control decisions on a majority of predictions, rather than a few frames. Our attack manages to trick the detector into not seeing a Stop sign in a majority of the tested video frames.\\

Despite these observations, we stress that a key step towards understanding the vulnerability of the broad class of object detection models to physical adversarial examples is to create algorithms that can attack state-of-the-art object detectors. In this work, we have shown how to can extend the existing \algshort algorithm with positional and rotational invariance to attack object detectors in relatively controlled settings.

\section{Conclusion}
Starting from an algorithm to generate robust physical perturbations for {\em classifiers}, we extend it with positional and rotational invariance to generate physical perturbations for state-of-the-art object {\em detectors}---a broader class of deep neural networks that are used in dynamic settings to detect and label objects within scenes. Object detectors are popular in cyber-physical systems such as autonomous vehicles. We experiment with the YOLO v2 object detector, showing that it is possible to physically perturb a Stop sign such that the detector ignores it. When presented with a video of the adversarial poster perturbation, YOLO failed to recognize the sign in 85.6\% of the video frames in a controlled lab environment, and in 72.5\% of the video frames in an outdoor environment. When presented with a video of the adversarial sticker perturbation, YOLO failed to recognize the sign in 85\% of the video frames in a controlled lab environment, and in 63.5\% of the video frames in an outdoor environment. We also observed limited blackbox transferability to the Faster-RCNN detector. The poster perturbation fooled Faster R-CNN  in 85.9\% of the video frames in a controlled lab environment, and in 40.2\% of the video frames in an outdoor environment. Our work, thus, takes steps towards developing a more informed understanding of the vulnerability of object detectors to physical adversarial examples.


\section*{Acknowledgements}
We thank the reviewers for their insightful feedback. This work was supported in part by NSF grants 1422211, 1565252, 1616575, 1646392, 1740897, Berkeley Deep Drive, the Center for Long-Term Cybersecurity, FORCES (which receives support from the NSF), the Hewlett Foundation, the MacArthur Foundation, a UM-SJTU grant, and the UW Tech Policy Lab. 

{\footnotesize \bibliographystyle{acm}
\bibliography{reference}}

\end{document}